\documentclass[12pt]{article}
\usepackage{amsmath,amssymb,amsthm,bm}
\usepackage{array}
\usepackage{caption,graphicx}
\usepackage[all]{xy}

\def\beq{\begin{equation}}
\def\eqn#1{\beq\label{#1}}
\def\eeq{\end{equation}}

\def\bb {\begin {eqnarray}}
\def\eqnn#1{\bb\label{#1}}
\def\ee {\end {eqnarray}}

\def\eqref#1{(\ref#1)}
\def\sha{{\textstyle{7\over2}}}

\def\ta{{\tilde\alpha}}

\def\nn{\nonumber}
\def\nt{\noindent}
\def\nl{\hfill\break}

\def\llr{\longrightarrow}
\def\({\left(}
\def\){\right)}
\def\eps{\epsilon}
 
\def\lra{\longrightarrow}

\def\ha{{\textstyle{1\over2}}}

  \def\tV{{\tilde V}}

\def\r{\rho}

\def\a{\alpha}
\def\b{\beta}

\def\vr{\vert}

\def\D{{\Delta}}

\def\bbr{{I\!\!R}}
\def\bbn{I\!\!N}
\def\bbc{{C\kern-8.5pt I}\,}
\def\bac{{C\kern-5.5pt I}}
\def\bab{{C\kern-4.5pt I}}

\def\L{\Lambda}

\def\rank{{\rm rank}}
\def\downcirc#1{\mathop{\circ}\limits_{#1}}
\def\riga{-\kern-4pt - \kern-4pt -}
\font\fat=cmsy10 scaled\magstep5

\def\Bbullet{\raise-3pt\hbox{\fat\char"0F}}

\def\ca{{\cal A}}  \def\cc{{\cal C}}
\def\cd{{\cal D}} \def\ce{{\cal E}} \def\cf{{\cal F}}
\def\cg{{\cal G}} \def\ch{{\cal H}} \def\ci{{\cal I}}
 \def\ck{{\cal K}} 
\def\cm{{\cal M}} \def\cn{{\cal N}} 
\def\cp{{\cal P}}  
 \def\ct{{\cal T}}

\def\ido{intertwining differential operator}
\def\idos{intertwining differential operators}

 \def\ha{{\textstyle{\frac{1}{2}}}}

\def\eps{\epsilon}

\def\ca{{\cal A}}
\def\nn{\nonumber}

\def\nl{\hfil\break}
\def\nt{\noindent}

\def\lra{\longleftrightarrow}

\input epsf.tex
\newcount\figno
\def\fig#1#2#3{
\par\begingroup\parindent=0pt\leftskip=1cm\rightskip=1cm\parindent=0pt
\baselineskip=11pt \global\advance\figno by 1 
\epsfxsize=#3 \centerline{\epsfbox{#2}} \vskip 12pt
#1\par
\endgroup\par}
\def\figlabel#1{\xdef#1{\the\figno}}
\def\encadremath#1{\vbox{\hrule\hbox{\vrule\kern8pt\vbox{\kern8pt
\hbox{$\displaystyle #1$}\kern8pt} \kern8pt\vrule}\hrule}}

\begin{document}
\begin{titlepage}
\begin{flushright}
CERN-PH-TH/2012-143
\end{flushright}
\vskip 1in
\begin{center}

{\Large{Invariant Differential Operators for Non-Compact}}
\vskip 0.2cm
{\Large{ Lie Groups: the   $Sp(n,\bbr)$  Case}}\footnote{To appear in the Proceedings of
 the 9. International Workshop {\it Lie Theory and Its Applications
in Physics} (LT-9), Varna, Bulgaria, June 2011.}

 \vskip 0.8cm
{\large V.K. Dobrev}
\vskip 0.5cm
{\small Theory Division, Department of Physics, CERN,\\ CH-1211 Geneva 23, Switzerland} \\
{\small permanent address:} \\
{\small Institute for Nuclear Research and Nuclear Energy,}\\
{\small Bulgarian Academy of Sciences,}
\\ {\small Tsarigradsko Chaussee 72, BG-1784 Sofia, Bulgaria}

 \end{center}

\vskip 0.8cm

\begin{abstract}In the present paper we continue the project of systematic
construction of invariant differential operators on the example of
the non-compact  algebras  $sp(n,\bbr)$, in detail for ~$n=6$. Our
choice of these algebras is motivated by the fact that   they belong
to a narrow class of algebras, which we call 'conformal Lie
algebras', which have very similar properties to the conformal
algebras of  Minkowski space-time. We give the main multiplets and the main
 reduced multiplets of indecomposable elementary representations for ~$n=6$, including
the necessary data for all relevant invariant differential
operators. In fact, this gives by reduction also the cases for
~$n<6$, since   the main multiplet for fixed ~$n$~ coincides with
one reduced case for ~$n+1$.

\end{abstract}

\end{titlepage}
\vfill\eject

\section{Introduction}

Consider a Lie group ~$G$, e.g., the Lorentz, Poincar\'e, conformal
groups, and differential equations
$$ \ci~f ~~=~~ j $$
which are $G$-invariant. These
play a very important role in the description of physical
symmetries - recall, e.g., the early examples of Dirac, Maxwell, d'Allembert,
equations and nowadays the latest applications
of (super-)differential operators in conformal field theory,
supergravity, string theory, (for a recent review, cf. e.g.,
\cite{Ter}). Naturally, it is important to construct systematically
such invariant equations and operators.

In a recent paper \cite{Dobinv} we started the systematic explicit
construction of invariant differential operators. We gave an
explicit description of the building blocks, namely, the parabolic
subgroups and subalgebras from which the necessary representations
are induced. Thus we have set the stage for study of different
non-compact groups.

In the present paper we  focus on the groups ~$Sp(n,\bbr)$,
which are very interesting for several reasons. First of all, they
belong to the class of Hermitian symmetric spaces, i.e., the pair
$(G,K)$ is a Hermitian symmetric pair ($K$ is the maximal compact
subgroup of the noncompact semisimple group $G$). Further,
~$Sp(n,\bbr)$ ~  belong to a narrower class of groups/algebras,
which we call 'conformal Lie groups or algebras' since they have
very similar properties to the canonical conformal algebras
$so(n,2)$ of $n$-dimensional Minkowski space-time.  This class was
identified from our point of view in \cite{Dobeseven}. Besides
$so(n,2)$ it includes the algebras ~$su(n,n)$, ~$sp(n,\bbr)$,
~$so^*(4n)$, $E_{7(-25)}\,$, (omitting to mention coincidences
between the low-dimensional cases, cf. \cite{Dobeseven}).
The corresponding groups are also called Hermitian symmetric spaces of tube type
\cite{FaKo}.
  The same class was identified from different considerations in \cite{Guna}, where these groups/algebras
 were called 'conformal groups  of simple Jordan algebras'.  It was identified
 from still different considerations also in \cite{Mackder}, where
the objects of the class were called simple space-time symmetries
generalizing conformal symmetry.

In our further plans it shall be very useful that (as in
\cite{Dobinv}) we  follow a procedure in representation theory in
which \idos\ appear canonically \cite{Dob} and which procedure has
been generalized to the supersymmetry setting and to quantum groups.

The present paper is organized a follows. In section 2 we give the
preliminaries, actually recalling and adapting facts from
\cite{Dobinv}. In Section 3 we specialize to the ~$sp(n,\bbr)$~
case. In Section 4 we present some results on the multiplet
classification of the representations and \idos\ between them.

\section{Preliminaries}

 Let $G$ be a semisimple non-compact Lie group, and $K$ a
maximal compact subgroup of $G$. Then we have an Iwasawa
decomposition ~$G=KA_0N_0$, where ~$A_0$~ is abelian simply
connected vector subgroup of ~$G$, ~$N_0$~ is a nilpotent simply
connected subgroup of ~$G$~ preserved by the action of ~$A_0$.
Further, let $M_0$ be the centralizer of $A_0$ in $K$. Then the
subgroup ~$P_0 ~=~ M_0 A_0 N_0$~ is a minimal parabolic subgroup of
$G$. A parabolic subgroup ~$P ~=~ M' A' N'$~ is any subgroup of $G$
(including $G$ itself) which contains a minimal parabolic subgroup.

The importance of the parabolic subgroups comes from the fact that
the representations induced from them generate all (admissible)
irreducible representations of $G$ \cite{Lan}. For the
classification of all irreducible representations it is enough to
use only the so-called {\it cuspidal} parabolic subgroups
~$P=M'A'N'$, singled out by the condition that ~rank$\, M' =$
rank$\, M'\cap K$ \cite{Zhea,KnZu}, so that $M'$ has discrete series
representations \cite{Har}. However, often induction from
non-cuspidal parabolics is also convenient, cf.
\cite{EHW,Dobinv,Dobpeds,GPW}.

Let ~$\nu$~ be a (non-unitary) character of ~$A'$, ~$\nu\in\ca'^*$,
let ~$\mu$~ fix an irreducible representation ~$D^\mu$~ of ~$M'$~ on
a vector space ~$V_\mu\,$.

 We call the induced
representation ~$\chi =$ Ind$^G_{P}(\mu\otimes\nu \otimes 1)$~ an
~{\it elementary representation} of $G$ \cite{DMPPT}. (These are
called {\it generalized principal series representations} (or {\it
limits thereof}) in \cite{Knapp}.) Their spaces of functions are:
\eqn{fun} \cc_\chi ~=~ \{ \cf \in C^\infty(G,V_\mu) ~ \vr ~ \cf
(gman) ~=~ e^{-\nu(H)} \cdot D^\mu(m^{-1})\, \cf (g) \} \eeq where
~$a= \exp(H)\in A'$, ~$H\in\ca'\,$, ~$m\in M'$, ~$n\in N'$. The
representation action is the $left$ regular action: \eqn{lrr}
(\ct^\chi(g)\cf) (g') ~=~ \cf (g^{-1}g') ~, \quad g,g'\in G\ .\eeq

For our purposes we need to restrict to ~{\it maximal}~ parabolic
subgroups ~$P$, (so that $\rank\,A'=1$), that may not be cuspidal.
For the representations that we consider the character ~$\nu$~ is
parameterized by a real number ~$d$, called the conformal weight or
energy.

Further, let ~$\mu$~ fix a discrete series representation ~$D^\mu$~
of $M'$ on the Hilbert space ~$V_\mu\,$, or the so-called limit of a
discrete series representation (cf. \cite{Knapp}). Actually, instead
of the discrete series we can use the finite-dimensional
(non-unitary) representation of $M'$ with the same Casimirs.

An important ingredient in our considerations are the ~{\it
highest/lowest weight representations}~ of ~$\cg$. These can be
realized as (factor-modules of) Verma modules ~$V^\L$~ over
~$\cg^\bac$, where ~$\L\in (\ch^\bac)^*$, ~$\ch^\bac$ is a Cartan
subalgebra of ~$\cg^\bac$, weight ~$\L = \L(\chi)$~ is determined
uniquely from $\chi$ \cite{Dob}. In this setting we can consider
also unitarity, which here means positivity w.r.t. the Shapovalov
form in which the conjugation is the one singling out $\cg$ from
$\cg^\bac$.

Actually, since our ERs may be induced from finite-dimensional
representations of ~$\cm'$~ (or their limits) the Verma modules are
always reducible. Thus, it is more convenient to use ~{\it
generalized Verma modules} ~$\tV^\L$~ such that the role of the
highest/lowest weight vector $v_0$ is taken by the
(finite-dimensional) space ~$V_\mu\,v_0\,$. For the generalized
Verma modules (GVMs) the reducibility is controlled only by the
value of the conformal weight $d$. Relatedly, for the \idos{} only
the reducibility w.r.t. non-compact roots is essential.

One main ingredient of our approach is as follows. We group the
(reducible) ERs with the same Casimirs in sets called ~{\it
multiplets} \cite{Dobmul,Dob}. The multiplet corresponding to fixed
values of the Casimirs may be depicted as a connected graph, the
vertices of which correspond to the reducible ERs and the lines
between the vertices correspond to intertwining operators. The
explicit parametrization of the multiplets and of their ERs is
important for understanding of the situation.

In fact, the multiplets contain explicitly all the data necessary to
construct the \idos{}. Actually, the data for each \ido{} consists
of the pair ~$(\b,m)$, where $\b$ is a (non-compact) positive root
of ~$\cg^\bac$, ~$m\in\bbn$, such that the BGG \cite{BGG} Verma
module reducibility condition (for highest weight modules) is
fulfilled: \eqn{bggr} (\L+\r, \b^\vee ) ~=~ m \ , \quad \b^\vee
\equiv 2 \b /(\b,\b) \ .\eeq When (\ref{bggr}) holds then the Verma
module with shifted weight ~$V^{\L-m\b}$ (or ~$\tV^{\L-m\b}$ ~ for
GVM and $\b$ non-compact) is embedded in the Verma module ~$V^{\L}$
(or ~$\tV^{\L}$). This embedding is realized by a singular vector
~$v_s$~ determined by a polynomial ~$\cp_{m,\b}(\cg^-)$~ in the
universal enveloping algebra ~$(U(\cg_-))\ v_0\,$, ~$\cg^-$~ is the
subalgebra of ~$\cg^\bac$ generated by the negative root generators
\cite{Dix}.
 More explicitly, \cite{Dob}, ~$v^s_{m,\b} = \cp_{m,\b}\, v_0$ (or ~$v^s_{m,\b} = \cp_{m,\b}\, V_\mu\,v_0$ for GVMs).\footnote{For
explicit expressions for singular vectors we refer to
\cite{Dobsin}.} Then there exists \cite{Dob} an \ido{} \eqn{lido}
\cd_{m,\b} ~:~ \cc_{\chi(\L)} ~\llr ~ \cc_{\chi(\L-m\b)} \eeq given
explicitly by: \eqn{mido}\cd_{m,\b} ~=~ \cp_{m,\b}(\widehat{\cg^-})
\eeq where ~$\widehat{\cg^-}$~ denotes the $right$ action on the
functions ~$\cf$, cf. (\ref{fun}).

\section{The Non-Compact Lie Algebras $sp(n,\bbr)$}

\nt Let ~$n\geq 2$. Let ~$\cg ~=~ sp(n,\bbr)$, the split real form
of ~$sp(n,\bbc)=\cg^\bac$. The maximal compact subgroup of ~$\cg$~
is ~$\ck \cong u(1)\oplus su(n)$, ~$\dim_\bbr\,\cp = n(n+1)$,
~$\dim_\bbr\,\cn = n^2$. This algebra has discrete series
representations and highest/lowest weight representations.

The split rank is equal to $n$, while ~$\cm =0$.

The Satake diagram \cite{Sata} of ~$sp(n,\bbr)$~ is the same as the
Dynkin diagram of $sp(n,\bbc)$~:
$$\downcirc{{\a_1}}
\riga\downcirc{{\a_2}} \riga \dots \riga\downcirc{{\a_{n-1}}}
\Longleftarrow \downcirc{{\a_n}}$$ Also the root systems coincide.

We choose a ~{\it maximal} parabolic ~$\cp=\cm'\ca'\cn'$~ such that
~$\ca'\cong so(1,1)$, while the factor ~$\cm'$~ has the same
finite-dimensional (nonunitary) representations as the
finite-dimensional (unitary) representations of  the semi-simple
subalgebra of   ~$\ck$, i.e., ~$\cm' = ~sl(n,\bbr)$, cf.
\cite{Dobinv}. Thus, these induced representations are
representations of finite $\ck$-type \cite{HC}. Relatedly, the
number of ERs in the corresponding multiplets is equal to ~$\vr
W(\cg^\bac,\ch^\bac)\vr\, /\, \vr W(\ck^\bac,\ch^\bac)\vr ~=~ 2^n$,
cf. \cite{Bourb}, where ~$\ch$~ is a Cartan subalgebra of both
~$\cg$~ and ~$\ck$. Note also that ~$\ck^\bac \cong u(1)^\bac \oplus
sl(n,\bbc) \cong \cm'^\bac \oplus \ca'^\bac$. Finally, note that
~$\dim_\bbr\,\cn' = n(n+1)/2$.

We label   the signature of the ERs of $\cg$   as follows:
\eqn{sgnd}  \chi ~=~ \{\, n_1\,, \ldots,\, n_{n-1}\, ;\, c\, \} \ ,
\qquad n_j \in \bbn\ , \quad c = d- (n+1)/2 \eeq where the last
entry of ~$\chi$~ labels the characters of $\ca'\,$, and the first
$n-1$ entries are labels of the finite-dimensional nonunitary irreps
of $\cm'\,$, (or of the finite-dimensional unitary irreps of
~$su(n)$).

 The reason to use the parameter ~$c$~ instead of ~$d$~ is that the
parametrization of the ERs in the multiplets is given in a simpler
way, as we shall see.

Below we shall use the following conjugation on the
finite-dimensional entries of the signature: \eqn{conu}
(n_1,\ldots,n_{n-1})^* ~\doteq~ (n_{n-1},\ldots,n_{1}) \eeq

The ERs in the multiplet are related also by intertwining integral
  operators. The  integral operators were introduced by
Knapp and Stein \cite{KnSt}. In fact, these operators are defined
for any ER, not only for the reducible ones, the general action
being: \eqnn{} & G_{KS} ~:~ \cc_\chi ~ \llr ~ \cc_{\chi'} \ ,\cr
&\chi ~=~ \{\, n_1,\ldots,n_{n-1} \,;\, c\, \} \ , \cr &\chi' ~=~
\{\, (n_1,\ldots,n_{n-1})^* \,;\, -c\, \}  \ee The above action on
the signatures is also called restricted Weyl reflection, since it
represents the nontrivial element of the 2-element restricted Weyl
group which arises canonically with every maximal parabolic
subalgebra.\footnote{Generically, the Knapp-Stein operators can be
normalized so that indeed ~$G_{KS} \circ G_{KS} = {\rm
Id}_{\cc_\chi}\,$. However, this usually fails exactly for the
reducible ERs that form the multiplets, cf., e.g., \cite{DMPPT}.}

Further, we need more explicitly the root system of the algebra
~$sp(n,F)$.

In terms of the orthonormal basis $\eps_i\,$, ~$i=1,\ldots,n$, the
 positive roots are given by \eqn{spnrpos} \D^+ = \{ \eps_i \pm \eps_j, ~1 \leq
i <j \leq n; ~2\eps_i, 1 \leq i \leq n\} ,  \eeq while the simple
roots are: \eqn{spnrsmp} \pi = \{\a_i = \eps_i - \eps_{i+1}, ~1 \leq
i \leq n - 1; \ \a_n = 2\eps_n\} \eeq With our choice of
normalization of
  the long roots  ~$2\eps_k$~ have
length 4, while the short roots ~$\eps_i \pm \eps_j$~ have length 2.

From these the compact roots are those that form (by restriction)
the root system of the semisimple part of ~$\ck^\bac$, the rest are
noncompact, i.e., \eqnn{spnrcnc}  {\rm compact:}&~~~ \a_{ij}
~\equiv~\eps_i  -\eps_j,\ , \quad 1 \leq i < j \leq n \ ,
  \cr {\rm noncompact:}&~~~ \b_{ij} ~\equiv~ \eps_i
+\eps_j,\ ,  ~~ 1 \leq i \leq j \leq n \  \ee Thus, the only
non-compact simple root is ~$\a_n=\b_{nn}\,$.

We adopt the following ordering of the roots: \eqn{ord} \begin{matrix}
\b_{11}  &&&&& &&&\cr \vee &&&&&&&&\cr \b_{12} &
> & \b_{22} &&&&&\cr \vee && \vee &&&&&&\cr \ldots & \ldots & \ldots
&\ldots & \ldots &&&&\cr \vee  &  & \vee  &  & \ldots & & \vee &&\cr
\b_{1n} &
> & \b_{2n} & > & \ldots & > & \b_{n-1,n} & > &\b_{nn}=\a_n \cr \vee
&  & \vee  &  & \ldots & & \vee &&\cr \a_{1n} & > & \a_{2n} & > &
\ldots & > & \a_{n-1,n} =\a_{n-1}&&\cr \vee  &  & \vee  &  & \ldots
& &  &&\cr \ldots & \ldots & \ldots &\ldots & \ldots &&&&\cr \vee &&
\vee &&&&&&\cr \a_{13} & > & \a_{23}=\a_2 &&&&&\cr \vee && &&&&&&\cr
\a_{12}=\a_1 &  &  &&&&&\cr \end{matrix}\eeq This ordering is lexicographical
adopting the ordering of the ~$\eps$~: \eqn{orde} \eps_1 > \cdots >
\eps_n \eeq

Further, we shall use the so-called Dynkin labels: \eqn{dynk} m_i
~\equiv~ (\L+\r,\a^\vee_i)  \ , \quad i=1,\ldots,n,\eeq where ~$\L =
\L(\chi)$, ~$\r$ is half the sum of the positive roots of
~$\cg^\bac$.

We shall use also   the so-called Harish-Chandra parameters:
\eqn{dynhc} m_\b \equiv (\L+\r, \b )\ ,\eeq where $\b$ is any
positive root of $\cg^\bac$. These parameters are redundant, since
they are expressed in terms of the Dynkin labels, however,   some
statements are best formulated in their terms. In particular, in the
case of the noncompact roots we have: \eqnn{hclab} m_{\b_{ij}} ~&=&~
\Big( \sum_{s=i}^n + \sum_{s=j}^n \Big)   m_s \ , \qquad i<j \ ,\cr
m_{\b_{ii}} ~&=&~  \sum_{s=i}^n    m_s  \ee

Now we can give the correspondence between the signatures $\chi$ and
the highest weight $\L$. The explicit connection is: \eqn{rela} n_i
= m_i \ , \quad  c ~=~ -\ha (m_\ta + m_n) ~=~ -\,\ha(   m_1+\cdots +
m_{n-1} + 2m_n  )\eeq where ~$\ta ~=~ \b_{11}$~ is the highest root.

There are several types of multiplets: the main type, (which
contains maximal number of ERs/GVMs, the finite-dimensional and the
discrete series representations), and some reduced types of
multiplets.

In the next Section we give the main type of multiplets and the main
reduced types for $sp(n,\bbr)$ for $n\leq 6$.

\section{Multiplets}

\nt The multiplets of the main type are in 1-to-1 correspondence
with the finite-dimensional irreps of ~$sp(n,\bbr)$, i.e., they will
be labelled by  the ~$n$~ positive Dynkin labels    ~$m_i\in\bbn$.
As we mentioned, each such multiplet contains ~$2^n$ ERs/GVMs.  It
is difficult
 to give explicitly the multiplets for general ~$n$.
Thus, we shall give explicitly the case ~$n=6$~ which can still be
represented and comprehended, and then show how to obtain the cases
$n<6$.

\subsection{sp(6,$\bbr$)}

\subsubsection{Main multiplets}

\nt The main multiplets $R^6$ contain $64 (=2^6)$  ERs/GVMs whose
signatures can be given in the following pair-wise manner:
\eqnn{tablsps} &\chi_0^\pm ~=&~ \{\, (
 m_1,
 m_2,
 m_3,
 m_4,
 m_5)^\pm\,;\,\pm\ha(m_\ta + m_6  )\,\}  \\
&\chi_a^\pm ~=&~ \{\, (
 m_1,
 m_2,
 m_3,
 m_4,
 m_5+2m_6)^\pm\,;\,\pm\ha m_{15}\,\} \cr
&\chi_b^\pm ~=&~ \{\, (
 m_1,
 m_2,
 m_3,
 m_{45},
 m_5+2m_6)^\pm\,;\,\pm\ha m_{14}\,\} \cr
&\chi_c^\pm ~=&~ \{\, (
 m_1,
 m_2,
 m_{34},
 m_{5},
 m_{45}+2m_6)^\pm\,;\,\pm\ha m_{13}\,\} \cr
&\chi_{c'}^\pm ~=&~ \{\, (
 m_1,
 m_2,
 m_3,
 m_{45}+2m_6,
 m_5)^\pm\,;\,\pm\ha m_{14}\,\} \cr
&\chi_d^\pm ~=&~ \{\, (
 m_1,
 m_{23},
 m_{4},
 m_{5},
 m_{35}+2m_6)^\pm\,;\,\pm\ha m_{12}\,\} \cr
 &\chi_{d'}^\pm ~=&~ \{\, (
 m_1,
 m_2,
 m_{34},
 m_{5}+2m_6,
 m_{45})^\pm\,;\,\pm\ha m_{13}\,\} \cr
 &\chi_e^\pm ~=&~ \{\, (
 m_{12},
 m_{3},
 m_{4},
 m_{5},
 m_{25}+2m_6)^\pm\,;\,\pm\ha m_{1}\,\} \cr
&\chi_{e'}^\pm ~=&~ \{\, (
 m_1,
 m_{23},
 m_{4},
 m_{5}+2m_6,
 m_{35})^\pm\,;\,\pm\ha m_{12}\,\} \cr
 &\chi_{e''}^\pm ~=&~ \{\, (
 m_1,
 m_2,
 m_{35},
 m_{5}+2m_6,
 m_{4})^\pm\,;\,\pm\ha m_{13}\,\} \cr
&\chi_f^\pm ~=&~ \{\, (
 m_{2},
 m_{3},
 m_{4},
 m_{5},
 m_{15}+2m_6)^\pm\,;\,\mp\ha m_{1}\,\} \cr
&\chi_{f'}^\pm ~=&~ \{\, (
 m_{12},
 m_{3},
 m_{4},
 m_{5}+2m_6,
 m_{25})^\pm\,;\,\pm\ha m_{1}\,\} \cr
&\chi_{f''}^\pm ~=&~ \{\, (
 m_{1},
 m_{23},
 m_{45},
 m_{5}+2m_6,
 m_{34})^\pm\,;\,\pm\ha m_{12}\,\} \cr
&\chi_{f'''}^\pm ~=&~ \{\, (
 m_1,
 m_2,
 m_{35}+2m_6,
 m_{5},
 m_{4})^\pm\,;\,\pm\ha m_{13}\,\} \cr
&\chi_g^\pm ~=&~ \{\, (
 m_{2},
 m_{3},
 m_{4},
 m_{5}+2m_6,
 m_{15})^\pm\,;\,\mp\ha m_{1}\,\} \cr
&\chi_{g'}^\pm ~=&~ \{\, (
 m_{12},
 m_{3},
 m_{45},
 m_{5}+2m_6,
 m_{24})^\pm\,;\,\pm\ha m_{1}\,\} \cr
&\chi_{g''}^\pm ~=&~ \{\, (
 m_1,
 m_{23},
 m_{45}+2m_6,
 m_{5},
 m_{34})^\pm\,;\,\pm\ha m_{12}\,\} \cr
 &\chi_h^\pm ~=&~ \{\, (
 m_{2},
 m_{3},
 m_{45},
 m_{5}+2m_6,
 m_{14})^\pm\,;\,\mp\ha m_{1}\,\} \cr
 &\chi_{h'}^\pm ~=&~ \{\, (
 m_{12},
 m_{3},
 m_{45}+2m_6,
 m_{5},
 m_{24})^\pm\,;\,\pm\ha m_{1}\,\} \cr
&\chi_{h''}^\pm ~=&~ \{\, (
 m_2,
 m_{3},
 m_{45}+2m_6,
 m_{5},
 m_{14})^\pm\,;\,\mp\ha m_{1}\,\} \cr
 &\chi_j^\pm ~=&~ \{\, (
 m_{2},
 m_{34},
 m_{5},
 m_{45}+2m_6,
 m_{13})^\pm\,;\,\mp\ha m_{1}\,\} \cr
 &\chi_{j'}^\pm ~=&~ \{\, (
 m_{12},
 m_{34},
 m_{5},
 m_{45}+2m_6,
 m_{23})^\pm\,;\,\pm\ha m_{1}\,\} \cr
 &\chi_{j''}^\pm ~=&~ \{\, (
 m_{1},
 m_{24},
 m_{5},
 m_{45}+2m_6,
 m_{3})^\pm\,;\,\pm\ha m_{12}\,\} \cr
 &\chi_k^\pm ~=&~ \{\, (
 m_{2},
 m_{34},
 m_{5}+2m_6,
 m_{45},
 m_{13})^\pm\,;\,\mp\ha m_{1}\,\} \cr
 &\chi_{k'}^\pm ~=&~ \{\, (
 m_{12},
 m_{34},
 m_{5}+2m_6,
 m_{45},
 m_{23})^\pm\,;\,\pm\ha m_{1}\,\} \cr
 &\chi_{k''}^\pm ~=&~ \{\, (
 m_{1},
 m_{24},
 m_{5}+2m_6,
 m_{45},
 m_{3})^\pm\,;\,\pm\ha m_{12}\,\} \cr
 &\chi_\ell^\pm ~=&~ \{\, (
 m_{2},
 m_{35},
 m_{5}+2m_6,
 m_{4},
 m_{13})^\pm\,;\,\mp\ha m_{1}\,\} \cr
 &\chi_{\ell'}^\pm ~=&~ \{\, (
 m_{12},
 m_{35},
 m_{5}+2m_6,
 m_{4},
 m_{23})^\pm\,;\,\pm\ha m_{1}\,\} \cr
 &\chi_{\ell''}^\pm ~=&~ \{\, (
 m_{1},
 m_{25},
 m_{5}+2m_6,
 m_{4},
 m_{3})^\pm\,;\,\pm\ha m_{12}\,\} \cr
&\chi_m^\pm ~=&~ \{\, (
 m_{2},
 m_{35}+2m_6,
 m_{5},
 m_{4},
 m_{13})^\pm\,;\,\mp\ha m_{1}\,\} \cr
 &\chi_{m'}^\pm ~=&~ \{\, (
 m_{12},
 m_{35}+2m_6,
 m_{5},
 m_{4},
 m_{23})^\pm\,;\,\pm\ha m_{1}\,\} \cr
 &\chi_{m''}^\pm ~=&~ \{\, (
 m_{1},
 m_{25}+2m_6,
 m_{5},
 m_{4},
 m_{3})^\pm\,;\,\pm\ha m_{12}\,\}  \nn\ee where the notation ~$(...)^\pm$~ employs the   conjugation
(\ref{conu})~: \eqn{conut}
  (n_1,...,n_{5})^- ~=~ (n_1,...,n_{5})\ , \qquad
(n_1,...,n_{5})^+ ~=~  (n_1,...,n_{5})^* ~=~ (n_5,...,n_{1})\eeq

Obviously, the pairs in (\ref{tablsps})  are related by Knapp-Stein
integral operators, i.e., \eqn{ackin}  G_{KS} ~:~ \cc_{\chi^\mp}
\lra \cc_{\chi^\pm} \eeq

Matters are arranged so that in every multiplet only the ER with
signature ~$\chi_0^-$~ contains a finite-dimensional nonunitary
subrepresentation in  a finite-dimensional subspace ~$\ce$. The
latter corresponds to the finite-dimensional   irrep of ~$sp(6)$~
with signature ~$\{ m_1\,,\ldots\,, m_6 \}$. The subspace ~$\ce$~ is
annihilated by the operator ~$G^+\,$,\ and is the image of the
operator ~$G^-\,$. The subspace ~$\ce$~ is annihilated also by the
\ido{} acting from ~$\chi^-$~ to ~$\chi'^-$~ (more about this
operator below).
 When all ~$m_i=1$~ then ~$\dim\,\ce = 1$, and in that case
~$\ce$~ is also the trivial one-dimensional UIR of the whole algebra
~$\cg$. Furthermore in that case the conformal weight is zero:
~$d=\sha+c=\sha-\ha(m_1+\cdots+m_5+2m_6)_{\vert_{m_i=1}}=0$.

Analogously, in every multiplet only the ER with signature
~$\chi_0^+$~ contains holomorphic discrete series representation.
This is guaranteed by the criterion \cite{Har} that for such an ER
all Harish-Chandra parameters for non-compact roots must be
negative, i.e., in our situation, ~$ m_\a ~<~ 0$, for ~$\a$~ from
the second row of (\ref{spnrcnc}). [That this holds for our
~$\chi^+$~ can be easily checked using the signatures
(\ref{tablsps}).]

In fact, the Harish-Chandra parameters are reflected in the division
of the ERs into ~$\chi^-$~ and ~$\chi^+$~: ~for the ~$\chi^-$~ less
than half of the 21 non-compact Harish-Chandra parameters are
negative, (none for ~$\chi_0^-$), while ~for the ~$\chi^+$~ more
than half of the 21 non-compact Harish-Chandra parameters are
negative, (all for ~$\chi_0^+$),

Note that the ER ~$\chi_0^+\,$~ contains also the conjugate
anti-holomorphic discrete series. The direct sum of the holomorphic
and the antiholomorphic representations are realized in  an
invariant subspace ~$\cd$~ of the ER ~$\chi_0^+\,$. That subspace is
annihilated by the operator ~$G^-\,$,\ and is the image of the
operator ~$G^+\,$.\nl Note that the corresponding lowest weight GVM
is infinitesimally equivalent only to the holomorphic discrete
series, while the conjugate highest weight GVM is infinitesimally
equivalent to the anti-holomorphic discrete series.\nl The conformal
weight of the ER  ~$\chi_0^+$~ has the restriction ~$d = \sha+c =
\sha + \ha(m_1+\cdots+m_5+2m_6)  \geq 7$.

The multiplets are given explicitly in Fig. 1, where we use the
notation: ~$\L^\pm = \L(\chi^\pm)$.  Each \ido\ is represented by an
arrow accompanied by a symbol ~$i_{j...k}$~ encoding the root
~$\b_{j...k}$~ and the number $m_{\b_{j...k}}$ which is involved in
the BGG criterion. This notation is used to save space, but it can
be used due to
 the fact that only \idos\ which are
non-composite are displayed, and that the data ~$\b,m_\b\,$, which
is involved in the embedding ~$V^\L \lra V^{\L-m_\b,\b}$~ turns out
to involve only the ~$m_i$~ corresponding to simple roots, i.e., for
each $\b,m_\b$ there exists ~$i = i(\b,m_\b,\L)\in \{
1,\ldots,2n-1\}$, such that ~$m_\b=m_i\,$. Hence the data
~$\b_{j...k}\,$,~$m_{\b_{j...k}}$~ is represented by ~$i_{j...k}$~
on the arrows.

The pairs ~$\L^\pm$~ are symmetric w.r.t. to the bullet in the
middle of the figure - this represents the Weyl symmetry realized by
the Knapp-Stein operators.

\subsubsection{Reduced multiplets $R^6_1$}

\nt The reduced multiplets of type $R^6_1$ contain 48 ERs/GVMs whose
signatures can be given in the following pair-wise manner:
\eqnn{tablone} &\chi_0^\pm ~=&~ \{\, (
 0,
 m_2,
 m_3,
 m_4,
 m_5)^\pm\,;\,\pm\ha(m_{25}  + 2m_6  )\,\}  \\
&\chi_a^\pm ~=&~ \{\, (
 0,
 m_2,
 m_3,
 m_4,
 m_5+2m_6)^\pm\,;\,\pm\ha m_{25}\,\} \cr
&\chi_b^\pm ~=&~ \{\, (
 0,
 m_2,
 m_3,
 m_{45},
 m_5+2m_6)^\pm\,;\,\pm\ha m_{24}\,\} \cr
&\chi_c^\pm ~=&~ \{\, (
 0,
 m_2,
 m_{34},
 m_{5},
 m_{45}+2m_6)^\pm\,;\,\pm\ha m_{23}\,\} \cr
&\chi_{c'}^\pm ~=&~ \{\, (
 0,
 m_2,
 m_3,
 m_{45}+2m_6,
 m_5)^\pm\,;\,\pm\ha m_{24}\,\} \cr
&\chi_d^\pm ~=&~ \{\, (
 0,
 m_{23},
 m_{4},
 m_{5},
 m_{35}+2m_6)^\pm\,;\,\pm\ha m_{2}\,\} \cr
 &\chi_{d'}^\pm ~=&~ \{\, (
 0,
 m_2,
 m_{34},
 m_{5}+2m_6,
 m_{45})^\pm\,;\,\pm\ha m_{23}\,\} \cr
 &\chi_e^\pm ~=&~ \{\, (
 m_{2},
 m_{3},
 m_{4},
 m_{5},
 m_{25}+2m_6)^\pm\,;\,0 \,\} \cr
&\chi_{e'}^\pm ~=&~ \{\, (
 0,
 m_{23},
 m_{4},
 m_{5}+2m_6,
 m_{35})^\pm\,;\,\pm\ha m_{2}\,\} \cr
 &\chi_{e''}^\pm ~=&~ \{\, (
 0,
 m_2,
 m_{35},
 m_{5}+2m_6,
 m_{4})^\pm\,;\,\pm\ha m_{23}\,\} \cr
&\chi_{f'}^\pm ~=&~ \{\, (
 m_{2},
 m_{3},
 m_{4},
 m_{5}+2m_6,
 m_{25})^\pm\,;\,0\,\} \cr
&\chi_{f''}^\pm ~=&~ \{\, (
 0,
 m_{23},
 m_{45},
 m_{5}+2m_6,
 m_{34})^\pm\,;\,\pm\ha m_{2}\,\} \cr
&\chi_{f'''}^\pm ~=&~ \{\, (
 0,
 m_2,
 m_{35}+2m_6,
 m_{5},
 m_{4})^\pm\,;\,\pm\ha m_{23}\,\} \cr
&\chi_{g'}^\pm ~=&~ \{\, (
 m_{2},
 m_{3},
 m_{45},
 m_{5}+2m_6,
 m_{24})^\pm\,;\,0\,\} \cr
&\chi_{g''}^\pm ~=&~ \{\, (
 0,
 m_{23},
 m_{45}+2m_6,
 m_{5},
 m_{34})^\pm\,;\,\pm\ha m_{2}\,\} \cr
 &\chi_{h'}^\pm ~=&~ \{\, (
 m_{2},
 m_{3},
 m_{45}+2m_6,
 m_{5},
 m_{24})^\pm\,;\,0\,\} \cr
 &\chi_{j'}^\pm ~=&~ \{\, (
 m_{2},
 m_{34},
 m_{5},
 m_{45}+2m_6,
 m_{23})^\pm\,;\,0\,\} \cr
 &\chi_{j''}^\pm ~=&~ \{\, (
 0,
 m_{24},
 m_{5},
 m_{45}+2m_6,
 m_{3})^\pm\,;\,\pm\ha m_{2}\,\} \cr
 &\chi_{k'}^\pm ~=&~ \{\, (
 m_{2},
 m_{34},
 m_{5}+2m_6,
 m_{45},
 m_{23})^\pm\,;\,0\,\} \cr
 &\chi_{k''}^\pm ~=&~ \{\, (
 0,
 m_{24},
 m_{5}+2m_6,
 m_{45},
 m_{3})^\pm\,;\,\pm\ha m_{2}\,\} \cr
 &\chi_{\ell'}^\pm ~=&~ \{\, (
 m_{2},
 m_{35},
 m_{5}+2m_6,
 m_{4},
 m_{23})^\pm\,;\,0\,\} \cr
 &\chi_{\ell''}^\pm ~=&~ \{\, (
 0,
 m_{25},
 m_{5}+2m_6,
 m_{4},
 m_{3})^\pm\,;\,\pm\ha m_{2}\,\} \cr
 &\chi_{m'}^\pm ~=&~ \{\, (
 m_{2},
 m_{35}+2m_6,
 m_{5},
 m_{4},
 m_{23})^\pm\,;\,0\,\} \cr
 &\chi_{m''}^\pm ~=&~ \{\, (
 0,
 m_{25}+2m_6,
 m_{5},
 m_{4},
 m_{3})^\pm\,;\,\pm\ha m_{2}\,\} \nn\ee
 The multiplets are given explicitly in Fig. 1a.

\subsubsection{Reduced multiplets $R^6_2$}

\nt The reduced multiplets of type $R^6_2$ contain 48 ERs/GVMs whose
signatures can be given in the following pair-wise manner:
\eqnn{tabltwo} &\chi_0^\pm ~=&~ \{\, (
 m_1,
 0,
 m_3,
 m_4,
 m_5)^\pm\,;\,\pm\ha(m_{1,35}  + 2m_6  )\,\}  \\
&\chi_a^\pm ~=&~ \{\, (
 m_1,
 0,
 m_3,
 m_4,
 m_5+2m_6)^\pm\,;\,\pm\ha m_{1,35}\,\} \cr
&\chi_b^\pm ~=&~ \{\, (
 m_1,
 0,
 m_3,
 m_{45},
 m_5+2m_6)^\pm\,;\,\pm\ha m_{1,34}\,\} \cr
&\chi_c^\pm ~=&~ \{\, (
 m_1,
 0,
 m_{34},
 m_{5},
 m_{45}+2m_6)^\pm\,;\,\pm\ha m_{1,3}\,\} \cr
&\chi_{c'}^\pm ~=&~ \{\, (
 m_1,
 0,
 m_3,
 m_{45}+2m_6,
 m_5)^\pm\,;\,\pm\ha m_{1,34}\,\} \cr
&\chi_d^\pm ~=&~ \{\, (
 m_1,
 m_{3},
 m_{4},
 m_{5},
 m_{35}+2m_6)^\pm\,;\,\pm\ha m_{1}\,\} \cr
 &\chi_{d'}^\pm ~=&~ \{\, (
 m_1,
 0,
 m_{34},
 m_{5}+2m_6,
 m_{45})^\pm\,;\,\pm\ha m_{1,3}\,\} \cr
&\chi_{e'}^\pm ~=&~ \{\, (
 m_1,
 m_{3},
 m_{4},
 m_{5}+2m_6,
 m_{35})^\pm\,;\,\pm\ha m_{1}\,\} \cr
 &\chi_{e''}^\pm ~=&
 ~ \{\, (
 m_1,
 0,
 m_{35},
 m_{5}+2m_6,
 m_{4})^\pm\,;\,\pm\ha m_{1,3}\,\} \cr
&\chi_f^\pm ~=&~ \{\, (
 0,
 m_{3},
 m_{4},
 m_{5},
 m_{1,35}+2m_6)^\pm\,;\,\mp\ha m_{1}\,\} \cr
&\chi_{f''}^\pm ~=&~ \{\, (
 m_{1},
 m_{3},
 m_{45},
 m_{5}+2m_6,
 m_{34})^\pm\,;\,\pm\ha m_{1}\,\} \cr
&\chi_{f'''}^\pm ~=&~ \{\, (
 m_1,
 0,
 m_{35}+2m_6,
 m_{5},
 m_{4})^\pm\,;\,\pm\ha m_{1,3}\,\} \cr
&\chi_g^\pm ~=&~ \{\, (
 0,
 m_{3},
 m_{4},
 m_{5}+2m_6,
 m_{1,35})^\pm\,;\,\mp\ha m_{1}\,\} \cr
&\chi_{g''}^\pm ~=&~ \{\, (
 m_1,
 m_{3},
 m_{45}+2m_6,
 m_{5},
 m_{34})^\pm\,;\,\pm\ha m_{1}\,\} \cr
 &\chi_h^\pm ~=&~ \{\, (
 0,
 m_{3},
 m_{45},
 m_{5}+2m_6,
 m_{1,34})^\pm\,;\,\mp\ha m_{1}\,\} \cr
&\chi_{h''}^\pm ~=&~ \{\, (
 0,
 m_{3},
 m_{45}+2m_6,
 m_{5},
 m_{1,34})^\pm\,;\,\mp\ha m_{1}\,\} \cr
 &\chi_j^\pm ~=&~ \{\, (
 0,
 m_{34},
 m_{5},
 m_{45}+2m_6,
 m_{1,3})^\pm\,;\,\mp\ha m_{1}\,\} \cr
 &\chi_{j''}^\pm ~=&~ \{\, (
 m_{1},
 m_{34},
 m_{5},
 m_{45}+2m_6,
 m_{3})^\pm\,;\,\pm\ha m_{1}\,\} \cr
 &\chi_k^\pm ~=&~ \{\, (
 0,
 m_{34},
 m_{5}+2m_6,
 m_{45},
 m_{1,3})^\pm\,;\,\mp\ha m_{1}\,\} \cr
 &\chi_{k''}^\pm ~=&~ \{\, (
 m_{1},
 m_{34},
 m_{5}+2m_6,
 m_{45},
 m_{3})^\pm\,;\,\pm\ha m_{1}\,\} \cr
 &\chi_\ell^\pm ~=&~ \{\, (
 0,
 m_{35},
 m_{5}+2m_6,
 m_{4},
 m_{1,3})^\pm\,;\,\mp\ha m_{1}\,\} \cr
 &\chi_{\ell''}^\pm ~=&~ \{\, (
 m_{1},
 m_{35},
 m_{5}+2m_6,
 m_{4},
 m_{3})^\pm\,;\,\pm\ha m_{1}\,\} \cr
 &\chi_m^\pm ~=&~ \{\, (
 0,
 m_{35}+2m_6,
 m_{5},
 m_{4},
 m_{1,3})^\pm\,;\,\mp\ha m_{1}\,\} \cr
 &\chi_{m''}^\pm ~=&~ \{\, (
 m_{1},
 m_{35}+2m_6,
 m_{5},
 m_{4},
 m_{3})^\pm\,;\,\pm\ha m_{1}\,\} \nn\ee
The multiplets are given explicitly in Fig. 1b.

\subsubsection{Reduced multiplets $R^6_3$}

\nt The reduced multiplets of type $R^6_3$ contain 48 ERs/GVMs whose
signatures can be given in the following pair-wise manner:
\eqnn{tablthree} &\chi_0^\pm ~=&~ \{\, (
 m_1,
 m_2,
 0,
 m_4,
 m_5)^\pm\,;\,\pm\ha(m_{12,45} + 2m_6  )\,\}  \\
&\chi_a^\pm ~=&~ \{\, (
 m_1,
 m_2,
 0,
 m_4,
 m_5+2m_6)^\pm\,;\,\pm\ha m_{12,45}\,\} \cr
&\chi_b^\pm ~=&~ \{\, (
 m_1,
 m_2,
 0,
 m_{45},
 m_5+2m_6)^\pm\,;\,\pm\ha m_{12,4}\,\} \cr
&\chi_{c'}^\pm ~=&~ \{\, (
 m_1,
 m_2,
 0,
 m_{45}+2m_6,
 m_5)^\pm\,;\,\pm\ha m_{12,4}\,\} \cr
&\chi_d^\pm ~=&~ \{\, (
 m_1,
 m_{2},
 m_{4},
 m_{5},
 m_{45}+2m_6)^\pm\,;\,\pm\ha m_{12}\,\} \cr
 &\chi_e^\pm ~=&~ \{\, (
 m_{12},
 0,
 m_{4},
 m_{5},
 m_{2,45}+2m_6)^\pm\,;\,\pm\ha m_{1}\,\} \cr
&\chi_{e'}^\pm ~=&~ \{\, (
 m_1,
 m_{2},
 m_{4},
 m_{5}+2m_6,
 m_{45})^\pm\,;\,\pm\ha m_{12}\,\} \cr
&\chi_f^\pm ~=&~ \{\, (
 m_{2},
 0,
 m_{4},
 m_{5},
 m_{12,45}+2m_6)^\pm\,;\,\mp\ha m_{1}\,\} \cr
&\chi_{f'}^\pm ~=&~ \{\, (
 m_{12},
 0,
 m_{4},
 m_{5}+2m_6,
 m_{2,45})^\pm\,;\,\pm\ha m_{1}\,\} \cr
&\chi_{f''}^\pm ~=&~ \{\, (
 m_{1},
 m_{2},
 m_{45},
 m_{5}+2m_6,
 m_{4})^\pm\,;\,\pm\ha m_{12}\,\} \cr
&\chi_g^\pm ~=&~ \{\, (
 m_{2},
 0,
 m_{4},
 m_{5}+2m_6,
 m_{12,45})^\pm\,;\,\mp\ha m_{1}\,\} \cr
&\chi_{g'}^\pm ~=&~ \{\, (
 m_{12},
 0,
 m_{45},
 m_{5}+2m_6,
 m_{2,4})^\pm\,;\,\pm\ha m_{1}\,\} \cr
&\chi_{g''}^\pm ~=&~ \{\, (
 m_1,
 m_{2},
 m_{45}+2m_6,
 m_{5},
 m_{4})^\pm\,;\,\pm\ha m_{12}\,\} \cr
 &\chi_h^\pm ~=&~ \{\, (
 m_{2},
 0,
 m_{45},
 m_{5}+2m_6,
 m_{12,4})^\pm\,;\,\mp\ha m_{1}\,\} \cr
 &\chi_{h'}^\pm ~=&~ \{\, (
 m_{12},
 0,
 m_{45}+2m_6,
 m_{5},
 m_{2,4})^\pm\,;\,\pm\ha m_{1}\,\} \cr
&\chi_{h''}^\pm ~=&~ \{\, (
 m_2,
 0,
 m_{45}+2m_6,
 m_{5},
 m_{12,4})^\pm\,;\,\mp\ha m_{1}\,\} \cr
 &\chi_j^\pm ~=&~ \{\, (
 m_{2},
 m_{4},
 m_{5},
 m_{45}+2m_6,
 m_{12})^\pm\,;\,\mp\ha m_{1}\,\} \cr
 &\chi_{j'}^\pm ~=&~ \{\, (
 m_{12},
 m_{4},
 m_{5},
 m_{45}+2m_6,
 m_{2})^\pm\,;\,\pm\ha m_{1}\,\} \cr
 &\chi_{j''}^\pm ~=&~ \{\, (
 m_{1},
 m_{2,4},
 m_{5},
 m_{45}+2m_6,
 0)^\pm\,;\,\pm\ha m_{12}\,\} \cr
 &\chi_k^\pm ~=&~ \{\, (
 m_{2},
 m_{4},
 m_{5}+2m_6,
 m_{45},
 m_{12})^\pm\,;\,\mp\ha m_{1}\,\} \cr
 &\chi_{k'}^\pm ~=&~ \{\, (
 m_{12},
 m_{4},
 m_{5}+2m_6,
 m_{45},
 m_{2})^\pm\,;\,\pm\ha m_{1}\,\} \cr
 &\chi_{k''}^\pm ~=&~ \{\, (
 m_{1},
 m_{2,4},
 m_{5}+2m_6,
 m_{45},
 0)^\pm\,;\,\pm\ha m_{12}\,\} \cr
 &\chi_\ell^\pm ~=&~ \{\, (
 m_{2},
 m_{45},
 m_{5}+2m_6,
 m_{4},
 m_{12})^\pm\,;\,\mp\ha m_{1}\,\} \cr
 &\chi_{\ell'}^\pm ~=&~ \{\, (
 m_{12},
 m_{45},
 m_{5}+2m_6,
 m_{4},
 m_{2})^\pm\,;\,\pm\ha m_{1}\,\} \cr
 &\chi_{\ell''}^\pm ~=&~ \{\, (
 m_{1},
 m_{2,45},
 m_{5}+2m_6,
 m_{4},
 0)^\pm\,;\,\pm\ha m_{12}\,\} \cr
&\chi_m^\pm ~=&~ \{\, (
 m_{2},
 m_{45}+2m_6,
 m_{5},
 m_{4},
 m_{12})^\pm\,;\,\mp\ha m_{1}\,\} \cr
 &\chi_{m'}^\pm ~=&~ \{\, (
 m_{12},
 m_{45}+2m_6,
 m_{5},
 m_{4},
 m_{2})^\pm\,;\,\pm\ha m_{1}\,\} \cr
 &\chi_{m''}^\pm ~=&~ \{\, (
 m_{1},
 m_{2,45}+2m_6,
 m_{5},
 m_{4},
 0)^\pm\,;\,\pm\ha m_{12}\,\}  \nn\ee
The multiplets are given explicitly in Fig. 1c.

\subsubsection{Reduced multiplets $R^6_4$}

\nt The reduced multiplets of type $R^6_4$ contain 48 ERs/GVMs whose
signatures can be given in the following pair-wise manner:
\eqnn{tablfour} &\chi_0^\pm ~=&~ \{\, (
 m_1,
 m_2,
 m_3,
 0,
 m_5)^\pm\,;\,\pm\ha(m_{13,5} + 2m_6  )\,\}  \\
&\chi_a^\pm ~=&~ \{\, (
 m_1,
 m_2,
 m_3,
 0,
 m_5+2m_6)^\pm\,;\,\pm\ha m_{13,5}\,\} \cr
&\chi_c^\pm ~=&~ \{\, (
 m_1,
 m_2,
 m_{3},
 m_{5},
 m_{5}+2m_6)^\pm\,;\,\pm\ha m_{13}\,\} \cr
&\chi_d^\pm ~=&~ \{\, (
 m_1,
 m_{23},
 0,
 m_{5},
 m_{3,5}+2m_6)^\pm\,;\,\pm\ha m_{12}\,\} \cr
 &\chi_{d'}^\pm ~=&~ \{\, (
 m_1,
 m_2,
 m_{3},
 m_{5}+2m_6,
 m_{5})^\pm\,;\,\pm\ha m_{13}\,\} \cr
 &\chi_e^\pm ~=&~ \{\, (
 m_{12},
 m_{3},
 0,
 m_{5},
 m_{23,5}+2m_6)^\pm\,;\,\pm\ha m_{1}\,\} \cr
&\chi_{e'}^\pm ~=&~ \{\, (
 m_1,
 m_{23},
 0,
 m_{5}+2m_6,
 m_{3,5})^\pm\,;\,\pm\ha m_{12}\,\} \cr
 &\chi_{e''}^\pm ~=&~ \{\, (
 m_1,
 m_2,
 m_{3,5},
 m_{5}+2m_6,
 0)^\pm\,;\,\pm\ha m_{13}\,\} \cr
&\chi_f^\pm ~=&~ \{\, (
 m_{2},
 m_{3},
 0,
 m_{5},
 m_{13,5}+2m_6)^\pm\,;\,\mp\ha m_{1}\,\} \cr
&\chi_{f'}^\pm ~=&~ \{\, (
 m_{12},
 m_{3},
 0,
 m_{5}+2m_6,
 m_{23,5})^\pm\,;\,\pm\ha m_{1}\,\} \cr
&\chi_{f'''}^\pm ~=&~ \{\, (
 m_1,
 m_2,
 m_{3,5}+2m_6,
 m_{5},
 0)^\pm\,;\,\pm\ha m_{13}\,\} \cr
&\chi_g^\pm ~=&~ \{\, (
 m_{2},
 m_{3},
 0,
 m_{5}+2m_6,
 m_{13,5})^\pm\,;\,\mp\ha m_{1}\,\} \cr
 &\chi_j^\pm ~=&~ \{\, (
 m_{2},
 m_{3},
 m_{5},
 m_{5}+2m_6,
 m_{13})^\pm\,;\,\mp\ha m_{1}\,\} \cr
 &\chi_{j'}^\pm ~=&~ \{\, (
 m_{12},
 m_{3},
 m_{5},
 m_{5}+2m_6,
 m_{23})^\pm\,;\,\pm\ha m_{1}\,\} \cr
 &\chi_{j''}^\pm ~=&~ \{\, (
 m_{1},
 m_{23},
 m_{5},
 m_{5}+2m_6,
 m_{3})^\pm\,;\,\pm\ha m_{12}\,\} \cr
 &\chi_k^\pm ~=&~ \{\, (
 m_{2},
 m_{3},
 m_{5}+2m_6,
 m_{5},
 m_{13})^\pm\,;\,\mp\ha m_{1}\,\} \cr
 &\chi_{k'}^\pm ~=&~ \{\, (
 m_{12},
 m_{3},
 m_{5}+2m_6,
 m_{5},
 m_{23})^\pm\,;\,\pm\ha m_{1}\,\} \cr
 &\chi_{k''}^\pm ~=&~ \{\, (
 m_{1},
 m_{23},
 m_{5}+2m_6,
 m_{5},
 m_{3})^\pm\,;\,\pm\ha m_{12}\,\} \cr
 &\chi_\ell^\pm ~=&~ \{\, (
 m_{2},
 m_{3,5},
 m_{5}+2m_6,
 0,
 m_{13})^\pm\,;\,\mp\ha m_{1}\,\} \cr
 &\chi_{\ell'}^\pm ~=&~ \{\, (
 m_{12},
 m_{3,5},
 m_{5}+2m_6,
 0,
 m_{23})^\pm\,;\,\pm\ha m_{1}\,\} \cr
 &\chi_{\ell''}^\pm ~=&~ \{\, (
 m_{1},
 m_{23,5},
 m_{5}+2m_6,
 0,
 m_{3})^\pm\,;\,\pm\ha m_{12}\,\} \cr
&\chi_m^\pm ~=&~ \{\, (
 m_{2},
 m_{3,5}+2m_6,
 m_{5},
 0,
 m_{13})^\pm\,;\,\mp\ha m_{1}\,\} \cr
 &\chi_{m'}^\pm ~=&~ \{\, (
 m_{12},
 m_{3,5}+2m_6,
 m_{5},
 0,
 m_{23})^\pm\,;\,\pm\ha m_{1}\,\} \cr
 &\chi_{m''}^\pm ~=&~ \{\, (
 m_{1},
 m_{23,5}+2m_6,
 m_{5},
 0,
 m_{3})^\pm\,;\,\pm\ha m_{12}\,\}  \nn\ee
The multiplets are given explicitly in Fig. 1d.

\subsubsection{Reduced multiplets $R^6_5$}

\nt The reduced multiplets of type $R^6_5$ contain 48 ERs/GVMs whose
signatures can be given in the following pair-wise manner:
\eqnn{tablfive} &\chi_0^\pm ~=&~ \{\, (
 m_1,
 m_2,
 m_3,
 m_4,
 0)^\pm\,;\,\pm\ha(m_{14}  + 2m_6  )\,\}  \\
&\chi_a^\pm ~=&~ \{\, (
 m_1,
 m_2,
 m_3,
 m_4,
 2m_6)^\pm\,;\,\pm\ha m_{14}\,\} \cr
&\chi_b^\pm ~=&~ \{\, (
 m_1,
 m_2,
 m_3,
 m_{4},
 2m_6)^\pm\,;\,\pm\ha m_{14}\,\} \cr
&\chi_c^\pm ~=&~ \{\, (
 m_1,
 m_2,
 m_{34},
 0,
 m_{4}+2m_6)^\pm\,;\,\pm\ha m_{13}\,\} \cr
&\chi_{c'}^\pm ~=&~ \{\, (
 m_1,
 m_2,
 m_3,
 m_{4}+2m_6,
 0)^\pm\,;\,\pm\ha m_{14}\,\} \cr
&\chi_d^\pm ~=&~ \{\, (
 m_1,
 m_{23},
 m_{4},
 0,
 m_{34}+2m_6)^\pm\,;\,\pm\ha m_{12}\,\} \cr
 &\chi_{d'}^\pm ~=&~ \{\, (
 m_1,
 m_2,
 m_{34},
 2m_6,
 m_{4})^\pm\,;\,\pm\ha m_{13}\,\} \cr
 &\chi_e^\pm ~=&~ \{\, (
 m_{12},
 m_{3},
 m_{4},
 0,
 m_{24}+2m_6)^\pm\,;\,\pm\ha m_{1}\,\} \cr
&\chi_{e'}^\pm ~=&~ \{\, (
 m_1,
 m_{23},
 m_{4},
 2m_6,
 m_{34})^\pm\,;\,\pm\ha m_{12}\,\} \cr
 &\chi_{e''}^\pm ~=&~ \{\, (
 m_1,
 m_2,
 m_{34},
 2m_6,
 m_{4})^\pm\,;\,\pm\ha m_{13}\,\} \cr
&\chi_f^\pm ~=&~ \{\, (
 m_{2},
 m_{3},
 m_{4},
 0,
 m_{14}+2m_6)^\pm\,;\,\mp\ha m_{1}\,\} \cr
&\chi_{f'}^\pm ~=&~ \{\, (
 m_{12},
 m_{3},
 m_{4},
 2m_6,
 m_{24})^\pm\,;\,\pm\ha m_{1}\,\} \cr
&\chi_{f''}^\pm ~=&~ \{\, (
 m_{1},
 m_{23},
 m_{4},
 2m_6,
 m_{34})^\pm\,;\,\pm\ha m_{12}\,\} \cr
&\chi_{f'''}^\pm ~=&~ \{\, (
 m_1,
 m_2,
 m_{34}+2m_6,
 0,
 m_{4})^\pm\,;\,\pm\ha m_{13}\,\} \cr
&\chi_g^\pm ~=&~ \{\, (
 m_{2},
 m_{3},
 m_{4},
 2m_6,
 m_{14})^\pm\,;\,\mp\ha m_{1}\,\} \cr
&\chi_{g'}^\pm ~=&~ \{\, (
 m_{12},
 m_{3},
 m_{4},
 2m_6,
 m_{24})^\pm\,;\,\pm\ha m_{1}\,\} \cr
&\chi_{g''}^\pm ~=&~ \{\, (
 m_1,
 m_{23},
 m_{4}+2m_6,
 0,
 m_{34})^\pm\,;\,\pm\ha m_{12}\,\} \cr
 &\chi_h^\pm ~=&~ \{\, (
 m_{2},
 m_{3},
 m_{4},
 2m_6,
 m_{14})^\pm\,;\,\mp\ha m_{1}\,\} \cr
 &\chi_{h'}^\pm ~=&~ \{\, (
 m_{12},
 m_{3},
 m_{4}+2m_6,
 0,
 m_{24})^\pm\,;\,\pm\ha m_{1}\,\} \cr
&\chi_{h''}^\pm ~=&~ \{\, (
 m_2,
 m_{3},
 m_{4}+2m_6,
 0,
 m_{14})^\pm\,;\,\mp\ha m_{1}\,\} \cr
 &\chi_j^\pm ~=&~ \{\, (
 m_{2},
 m_{34},
 0,
 m_{4}+2m_6,
 m_{13})^\pm\,;\,\mp\ha m_{1}\,\} \cr
 &\chi_{j'}^\pm ~=&~ \{\, (
 m_{12},
 m_{34},
 0,
 m_{4}+2m_6,
 m_{23})^\pm\,;\,\pm\ha m_{1}\,\} \cr
 &\chi_{j''}^\pm ~=&~ \{\, (
 m_{1},
 m_{24},
 0,
 m_{4}+2m_6,
 m_{3})^\pm\,;\,\pm\ha m_{12}\,\} \cr
 &\chi_k^\pm ~=&~ \{\, (
 m_{2},
 m_{34},
 2m_6,
 m_{4},
 m_{13})^\pm\,;\,\mp\ha m_{1}\,\} \cr
 &\chi_{k'}^\pm ~=&~ \{\, (
 m_{12},
 m_{34},
 2m_6,
 m_{4},
 m_{23})^\pm\,;\,\pm\ha m_{1}\,\} \cr
 &\chi_{k''}^\pm ~=&~ \{\, (
 m_{1},
 m_{24},
 2m_6,
 m_{4},
 m_{3})^\pm\,;\,\pm\ha m_{12}\,\} \cr
&\chi_m^\pm ~=&~ \{\, (
 m_{2},
 m_{34}+2m_6,
 0,
 m_{4},
 m_{13})^\pm\,;\,\mp\ha m_{1}\,\} \cr
 &\chi_{m'}^\pm ~=&~ \{\, (
 m_{12},
 m_{34}+2m_6,
 0,
 m_{4},
 m_{23})^\pm\,;\,\pm\ha m_{1}\,\} \cr
 &\chi_{m''}^\pm ~=&~ \{\, (
 m_{1},
 m_{24}+2m_6,
 0,
 m_{4},
 m_{3})^\pm\,;\,\pm\ha m_{12}\,\}  \nn\ee
The multiplets are given explicitly in Fig. 1e.

\subsubsection{Reduced multiplets $R^6_6$}

\nt The reduced multiplets of type $R^6_6$ contain 32 ERs/GVMs whose
signatures can be given in the following pair-wise manner:
\eqnn{tablsix} &\chi_0^\pm ~=&~ \{\, (
 m_1,
 m_2,
 m_3,
 m_4,
 m_5)^\pm\,;\,\pm\ha m_{15} \,\}  \\
&\chi_{c'}^\pm ~=&~ \{\, (
 m_1,
 m_2,
 m_3,
 m_{45},
 m_5)^\pm\,;\,\pm\ha m_{14}\,\} \cr
 &\chi_{d'}^\pm ~=&~ \{\, (
 m_1,
 m_2,
 m_{34},
 m_{5},
 m_{45})^\pm\,;\,\pm\ha m_{13}\,\} \cr
&\chi_{e'}^\pm ~=&~ \{\, (
 m_1,
 m_{23},
 m_{4},
 m_{5},
 m_{35})^\pm\,;\,\pm\ha m_{12}\,\} \cr
&\chi_{f'}^\pm ~=&~ \{\, (
 m_{12},
 m_{3},
 m_{4},
 m_{5},
 m_{25})^\pm\,;\,\pm\ha m_{1}\,\} \cr
&\chi_{f'''}^\pm ~=&~ \{\, (
 m_1,
 m_2,
 m_{35},
 m_{5},
 m_{4})^\pm\,;\,\pm\ha m_{13}\,\} \cr
&\chi_g^\pm ~=&~ \{\, (
 m_{2},
 m_{3},
 m_{4},
 m_{5},
 m_{15})^\pm\,;\,\mp\ha m_{1}\,\} \cr
&\chi_{g''}^\pm ~=&~ \{\, (
 m_1,
 m_{23},
 m_{45},
 m_{5},
 m_{34})^\pm\,;\,\pm\ha m_{12}\,\} \cr
 &\chi_{h'}^\pm ~=&~ \{\, (
 m_{12},
 m_{3},
 m_{45},
 m_{5},
 m_{24})^\pm\,;\,\pm\ha m_{1}\,\} \cr
&\chi_{h''}^\pm ~=&~ \{\, (
 m_2,
 m_{3},
 m_{45},
 m_{5},
 m_{14})^\pm\,;\,\mp\ha m_{1}\,\} \cr
 &\chi_k^\pm ~=&~ \{\, (
 m_{2},
 m_{34},
 m_{5},
 m_{45},
 m_{13})^\pm\,;\,\mp\ha m_{1}\,\} \cr
 &\chi_{k'}^\pm ~=&~ \{\, (
 m_{12},
 m_{34},
 m_{5},
 m_{45},
 m_{23})^\pm\,;\,\pm\ha m_{1}\,\} \cr
 &\chi_{k''}^\pm ~=&~ \{\, (
 m_{1},
 m_{24},
 m_{5},
 m_{45},
 m_{3})^\pm\,;\,\pm\ha m_{12}\,\} \cr
 &\chi_\ell^\pm ~=&~ \{\, (
 m_{2},
 m_{35},
 m_{5},
 m_{4},
 m_{13})^\pm\,;\,\mp\ha m_{1}\,\} \cr
 &\chi_{\ell'}^\pm ~=&~ \{\, (
 m_{12},
 m_{35},
 m_{5},
 m_{4},
 m_{23})^\pm\,;\,\pm\ha m_{1}\,\} \cr
 &\chi_{\ell''}^\pm ~=&~ \{\, (
 m_{1},
 m_{25},
 m_{5},
 m_{4},
 m_{3})^\pm\,;\,\pm\ha m_{12}\,\}\nn
\ee The multiplets are given explicitly in Fig. 1f.

Here the ER ~$\chi_0^+$~ contains the limits of the
(anti)holomorphic discrete series representations. This is
guaranteed by the fact that for this ER all Harish-Chandra
parameters for non-compact roots are non-positive, i.e., ~$ m_\a
~\leq~ 0$, for ~$\a$~ from  (\ref{hclab}). (Actually, we have:
~$m_{11}=0$, ~$ m_\a < 0$ for the rest of the non-compact $\a$.) Its
conformal weight has the restriction ~$d = \sha +
\ha(m_1+\cdots+m_5)  \geq 6$.

\subsection{The Cases sp(n,$\bbr$) for $n\leq 5$}

\nt We start with $sp(5,\bbr)$. The main multiplets $R^5$ contain
$32 (=2^5)$ ERs/GVMs whose signatures can be given in the following
pair-wise manner: \eqnn{tablfv} &\chi_0^\pm ~=&~ \{\, (
 m_1,
 m_2,
 m_3,
 m_4)^\pm\,;\,\pm\ha(m_{14} + 2m_5  )\,\}  \\
&\chi_a^\pm ~=&~ \{\, (
 m_1,
 m_2,
 m_3,
 m_4+2m_5)^\pm\,;\,\pm\ha m_{14}\,\} \cr
&\chi_b^\pm ~=&~ \{\, (
 m_1,
 m_2,
 m_{34},
 m_4+2m_5)^\pm\,;\,\pm\ha m_{13}\,\} \cr
&\chi_c^\pm ~=&~ \{\, (
 m_1,
 m_{23},
 m_{4},
 m_{34}+2m_5)^\pm\,;\,\pm\ha m_{12}\,\} \cr
&\chi_{c'}^\pm ~=&~ \{\, (
 m_1,
 m_2,
 m_{34}+2m_5,
 m_4)^\pm\,;\,\pm\ha m_{13}\,\} \cr
&\chi_d^\pm ~=&~ \{\, (
 m_{12},
 m_{3},
 m_{4},
 m_{24}+2m_5)^\pm\,;\,\pm\ha m_{1}\,\} \cr
 &\chi_{d'}^\pm ~=&~ \{\, (
 m_1,
 m_{23},
 m_{4}+2m_5,
 m_{34})^\pm\,;\,\pm\ha m_{12}\,\} \cr
 &\chi_e^\pm ~=&~ \{\, (
 m_{2},
 m_{3},
 m_{4},
 m_{14}+2m_5)^\pm\,;\,\mp\ha m_{1}\,\} \cr
&\chi_{e'}^\pm ~=&~ \{\, (
 m_{12},
 m_{3},
 m_{4}+2m_5,
 m_{24})^\pm\,;\,\pm\ha m_{1}\,\} \cr
 &\chi_{e''}^\pm ~=&~ \{\, (
 m_1,
 m_{24},
 m_{4}+2m_5,
 m_{3})^\pm\,;\,\pm\ha m_{12}\,\} \cr
&\chi_f^\pm ~=&~ \{\, (
 m_{2},
 m_{3},
 m_{4}+2m_5,
 m_{14}
 )^\pm\,;\,\mp\ha m_{1}\,\} \cr
&\chi_{f'}^\pm ~=&~ \{\, (
 m_{12},
 m_{34},
 m_{4}+2m_5,
 m_{23})^\pm\,;\,\pm\ha m_{1}\,\} \cr
&\chi_{f''}^\pm ~=&~ \{\, (
 m_{1},
m_{24}+2m_5,
 m_{4},
 m_{3})^\pm\,;\,\pm\ha m_{12}\,\} \cr
&\chi_g^\pm ~=&~ \{\, (
 m_{2},
 m_{34},
 m_{4}+2m_5,
 m_{13})^\pm\,;\,\mp\ha m_{1}\,\} \cr
&\chi_{g'}^\pm ~=&~ \{\, (
 m_{12},
m_{34}+2m_5,
 m_{4},
 m_{23})^\pm\,;\,\pm\ha m_{1}\,\} \cr
 &\chi_h^\pm ~=&~ \{\, (
 m_{2},
 m_{34}+2m_5,
 m_{4},
 m_{13})^\pm\,;\,\mp\ha m_{1}\,\} \nn \ee

Recalling that the ~$Sp(6,\bbr)$~ reduced multiplets of type
~$R^6_6$~ also have 32 members we check whether they may be
coinciding.  Indeed, that turns out to be the case and this is
obvious from the corresponding figures, Fig. 1f and Fig. 2 (though
our graphical representations  are a little distorted!). To make it
explicit via the signatures we do the following manipulations of
Table (\ref{tablsix})~: in each signature we just drop the entry
~$m_5$~ (there is exactly one such entry in each signature). Then we
replace each entry of the kind: ~$m_{k5}$, ($k=1,2,3,4$), by
~$m_{k4} + 2m_5$ (identifying ~$m_{44} \equiv m_4$). Thus
(\ref{tablsix}) becomes exactly (\ref{tablfv}). (Of course, this
does not mean that the contents is the same. For instance, the ER
~$\chi_0^+$~ from (\ref{tablfv}) contains
 the (anti)holomorphic discrete series representations
of $sp(5,\bbr)$, while the ER ~$\chi_0^+$~ from (\ref{tablsix})
contains the limits of the (anti)holomorphic discrete series
representations of $sp(6,\bbr)$.)

Thus, it is clear how to obtain from the case $sp(6,\bbr)$ all the
cases   $sp(n,\bbr)$ for $n\leq 5$.  We shall not do it here due to
the lack of space.

\section{Outlook}

 In the present paper we continued the programme outlined in
\cite{Dobinv} on the example of the non-compact group  $Sp(n,\bbr)$.
Similar explicit descriptions are planned for the other non-compact
groups, in particular, those  with  highest/lowest weight
representations. From the latter we have considered so far the cases
of $E_{7(-25)}$ \cite{Dobeseven}\footnote{For a different use of $E_{7(-25)}$, see,
e.g., \cite{CCM}.}, $E_{6(-14)}$ \cite{Dobesix},
$SU(n,n)$ ($n\leq4$) \cite{Dobsunn}. We plan also to extend these
considerations to the supersymmetric cases and also to the quantum
group setting. Such considerations are expected to be very useful
for applications to string theory and integrable models, cf., e.g.,
\cite{Witten}. In our further plans it shall be very useful that (as
in \cite{Dobinv}) we
 follow a procedure in representation
theory in which \idos\ appear canonically \cite{Dob} and which
procedure has been generalized to the supersymmetry setting
\cite{DobPet,Dobsusy} and to quantum groups \cite{Dobqg}.

\vskip 5mm

\noindent {\bf Acknowledgments.}
 This work was supported in part by the Bulgarian National Science Fund, grant DO 02-257. The author
thanks the Theory Division of CERN for hospitality during the course
of this work.

\fig{}{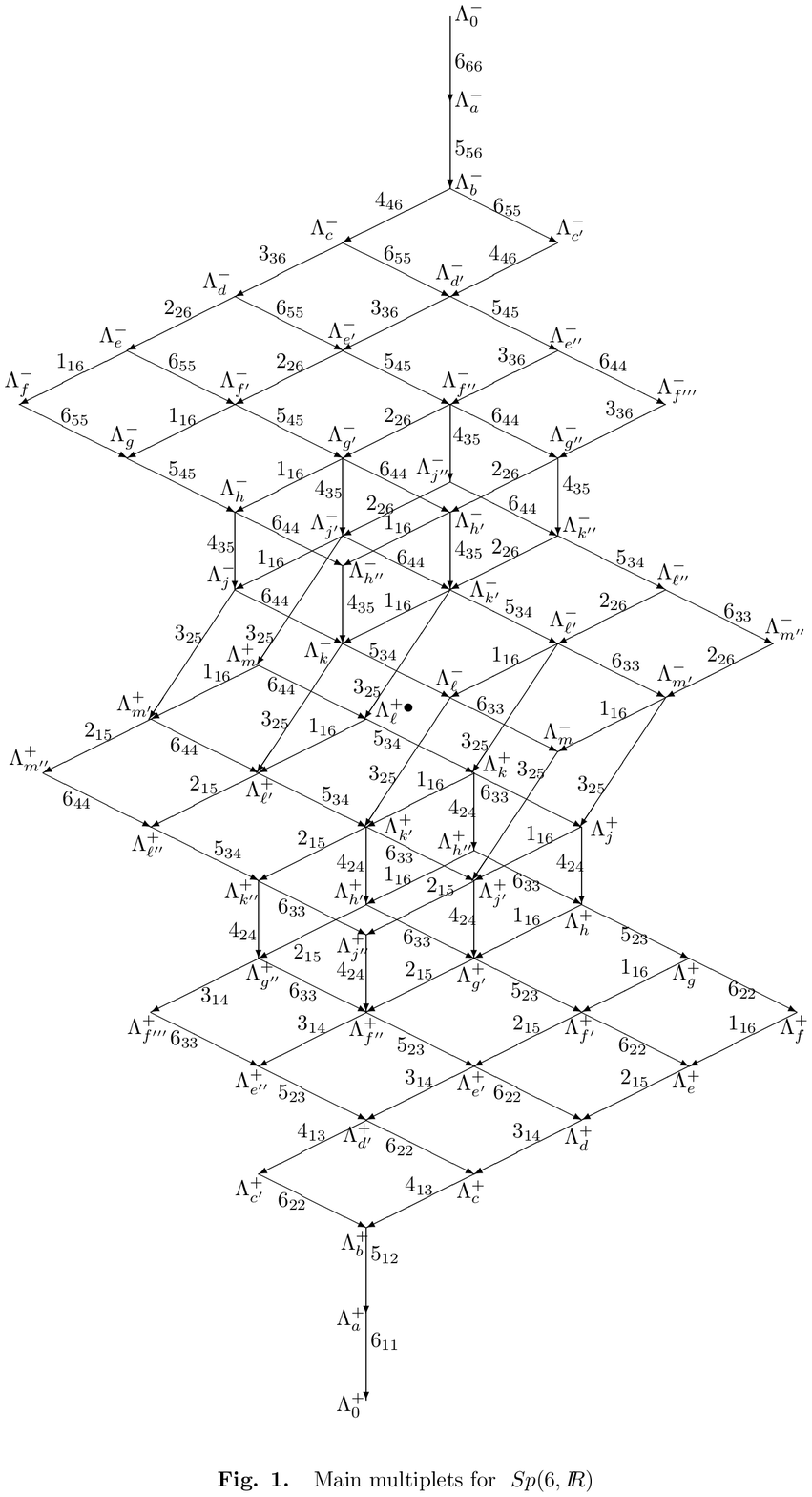}{160mm}

\fig{}{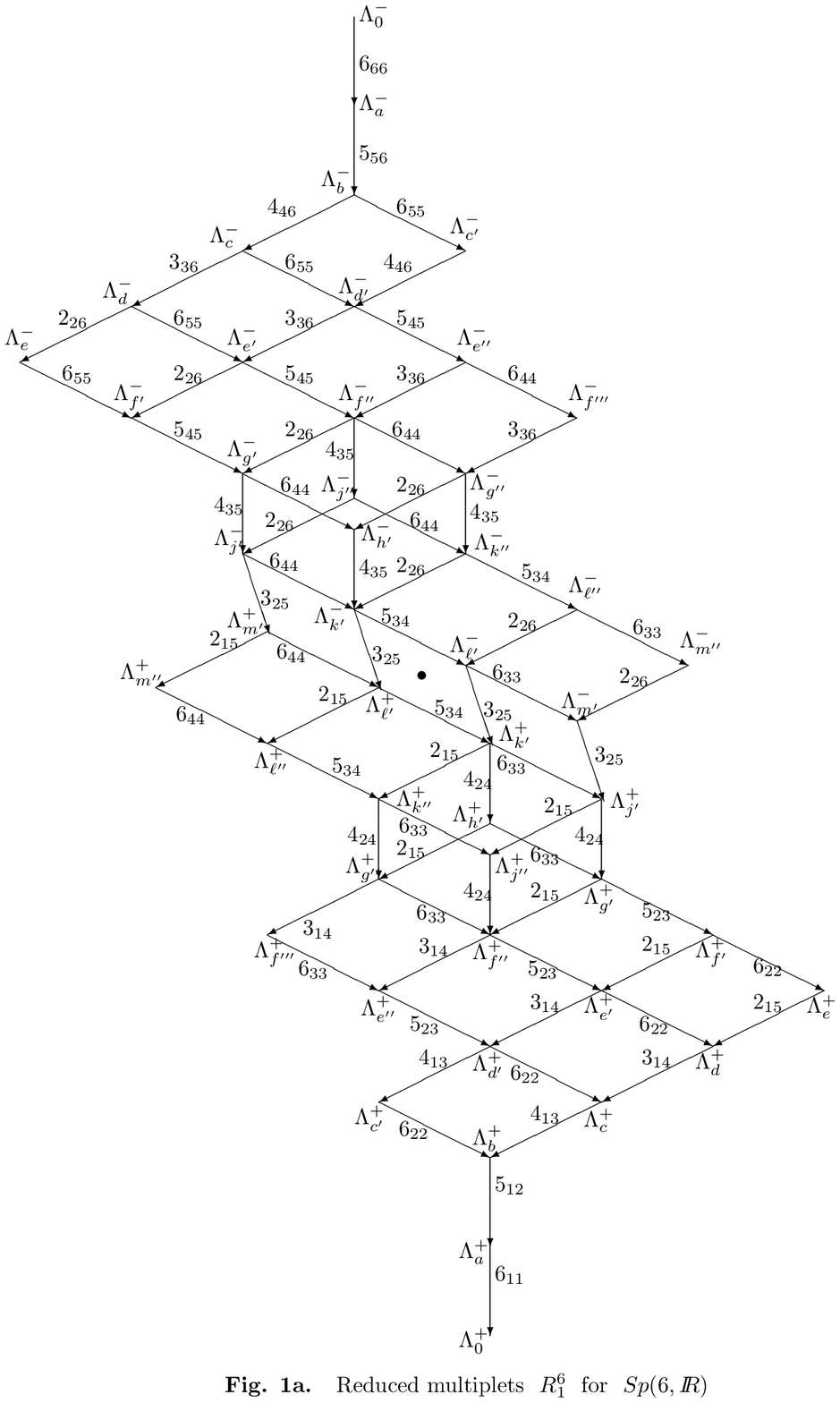}{115mm}

\fig{}{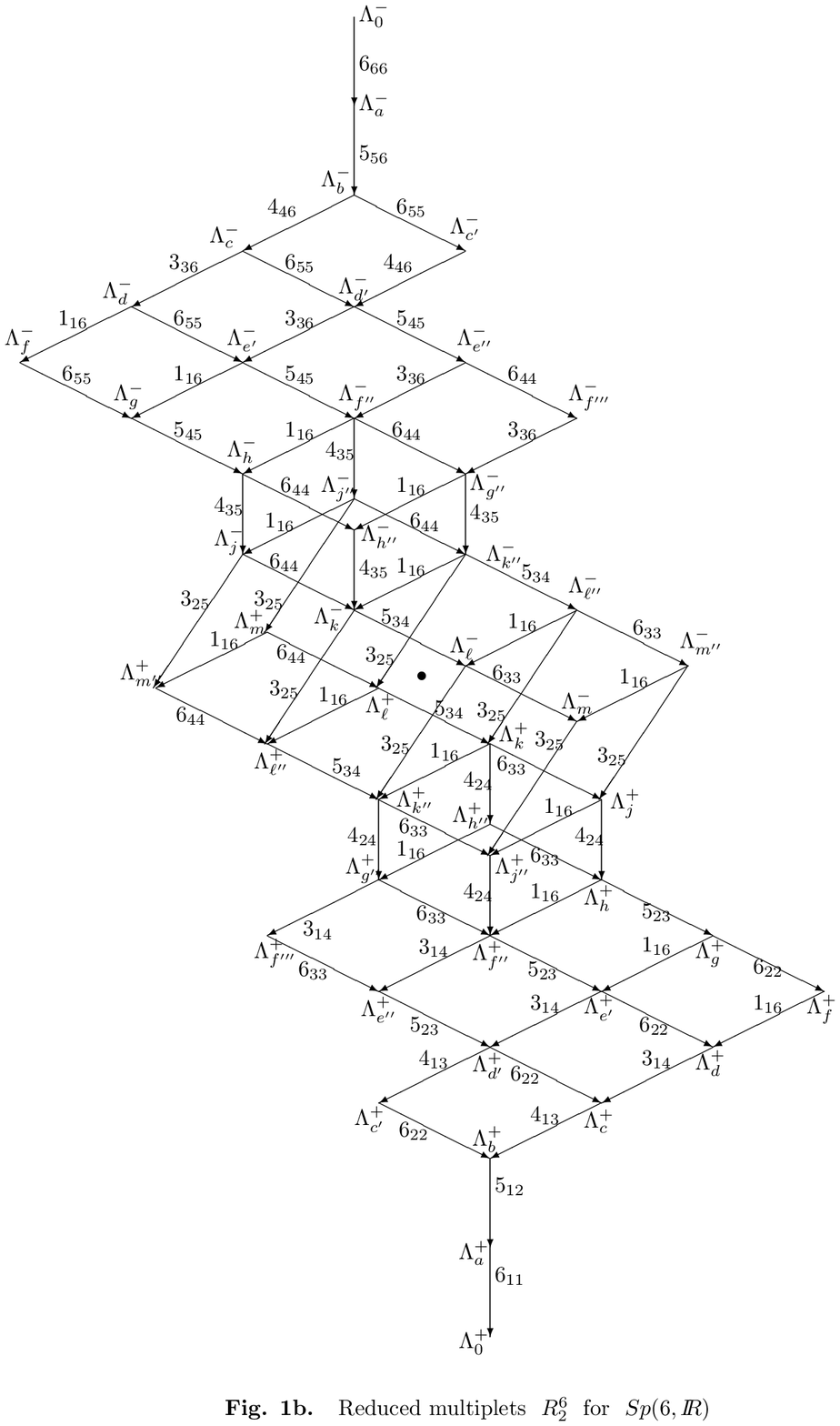}{115mm}

\fig{}{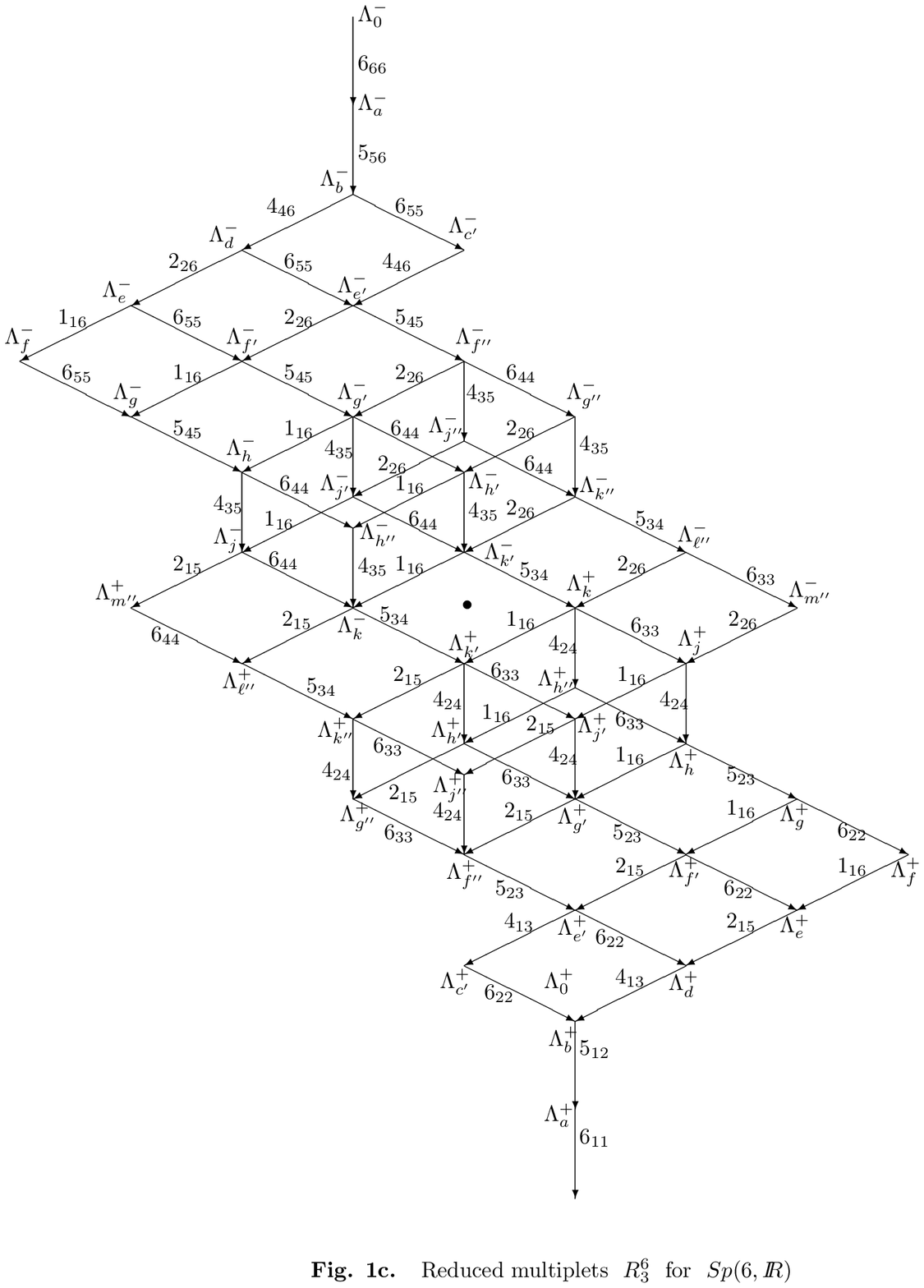}{14cm}

\fig{}{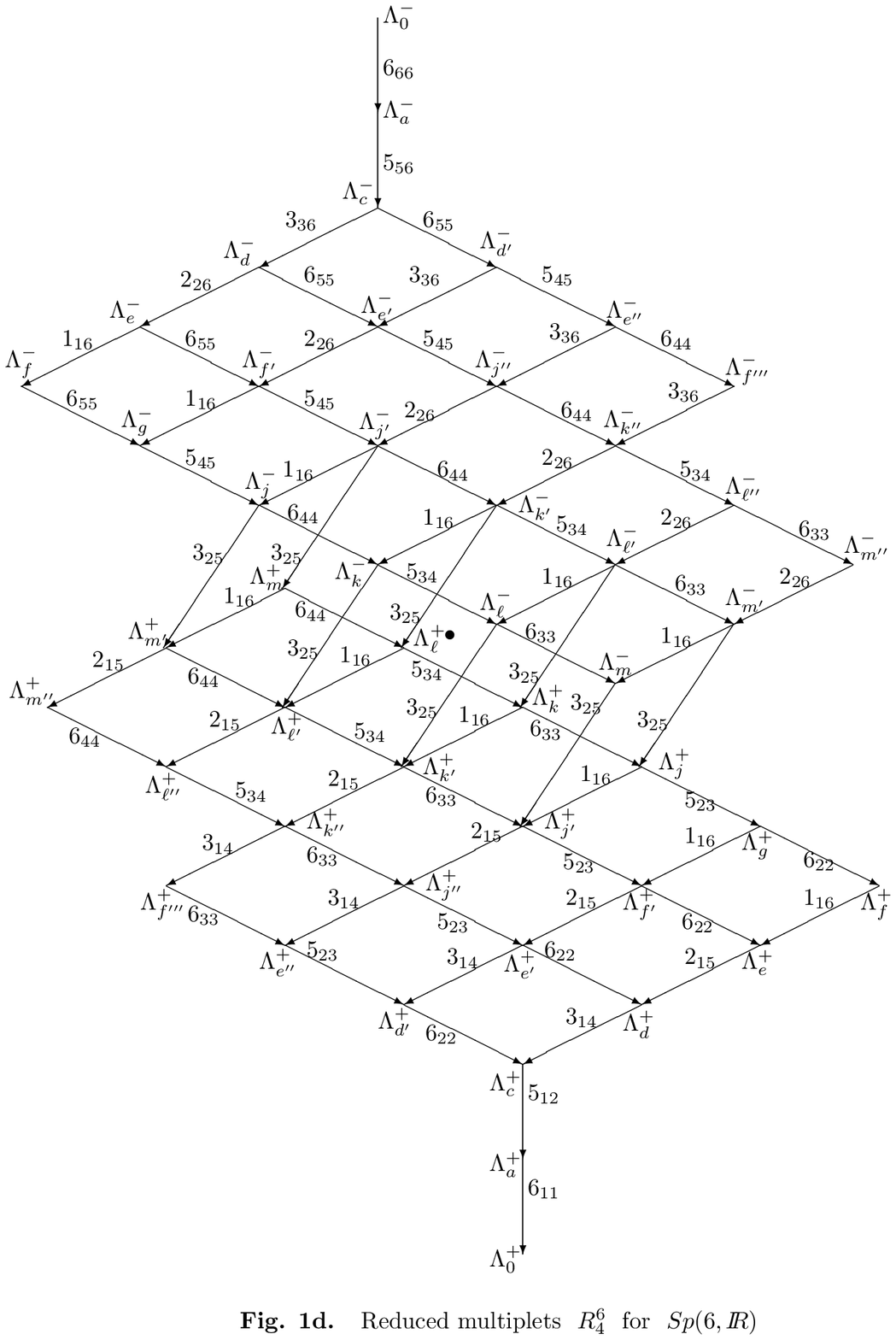}{130mm}

\fig{}{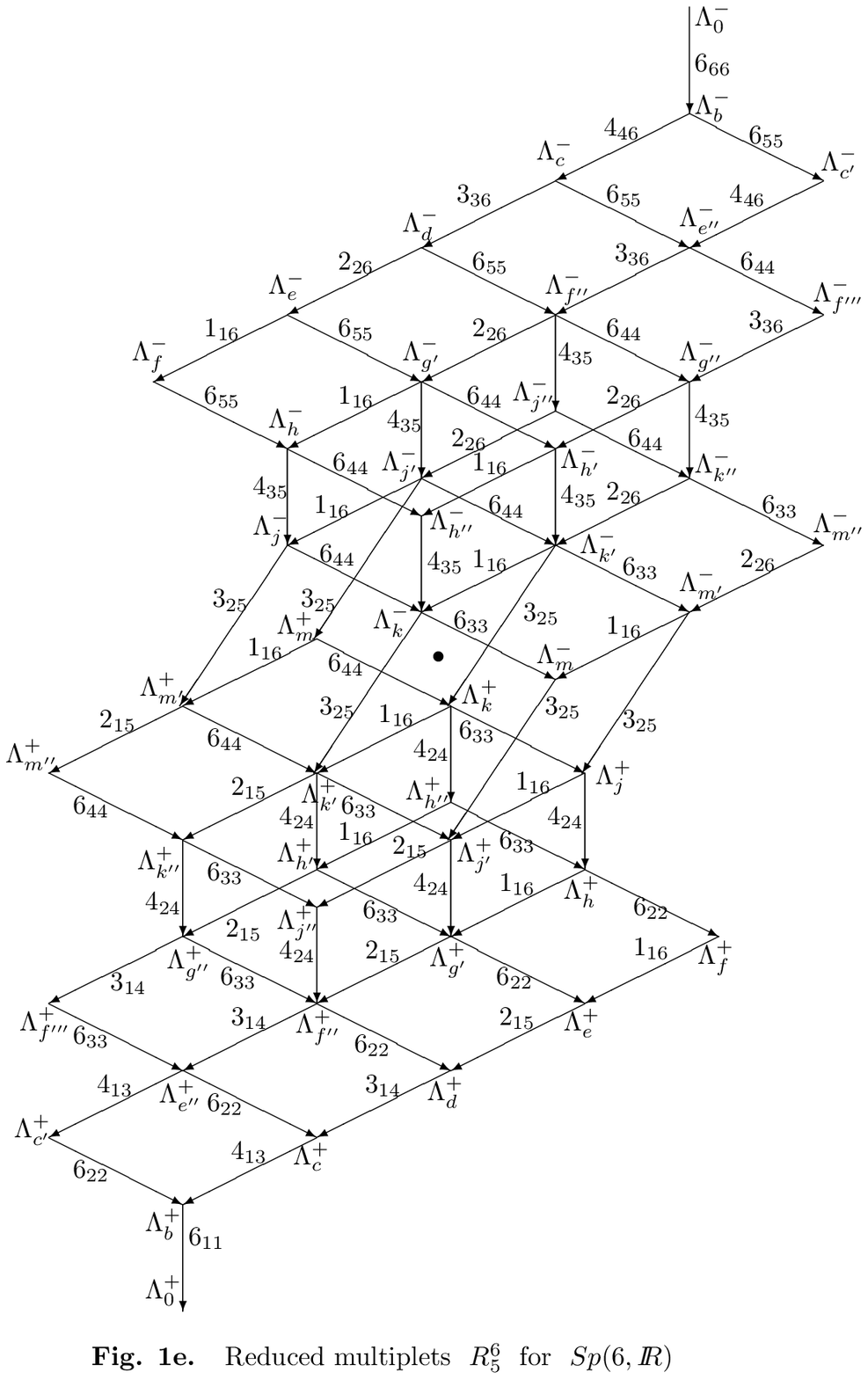}{120mm}

\fig{}{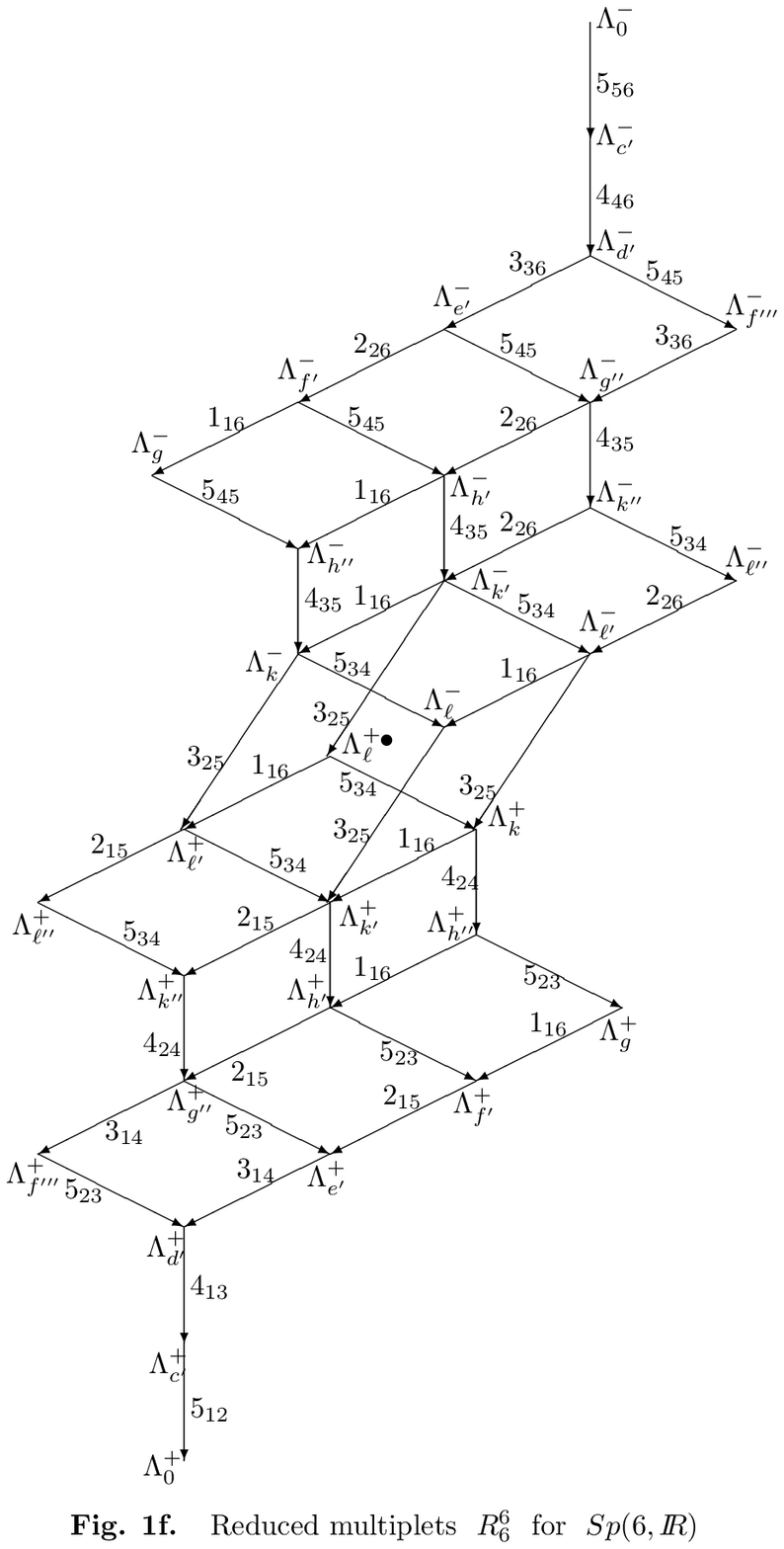}{95mm}

\fig{}{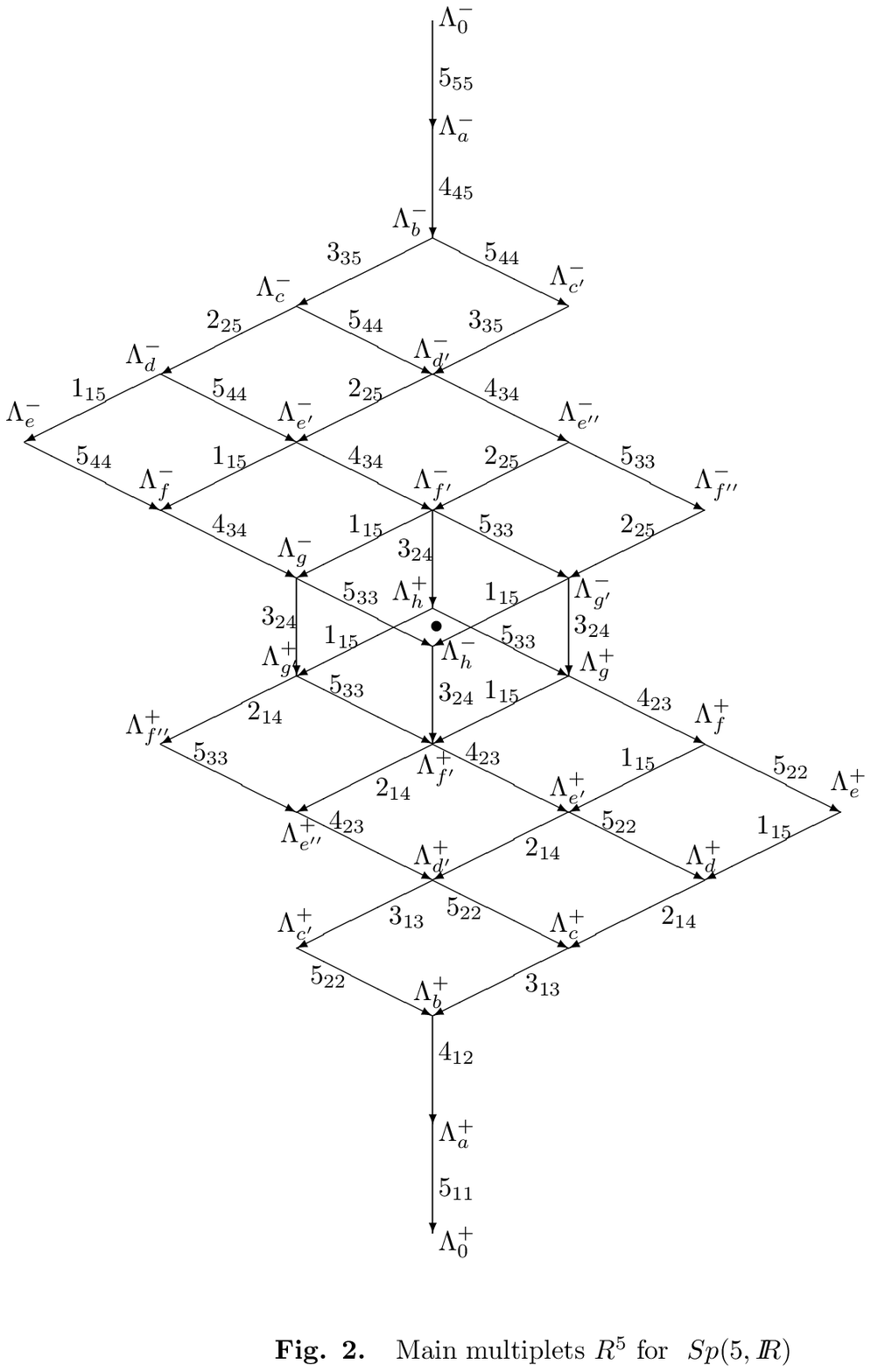}{120mm}

\end{document}